# Approximate Analytical Solutions of the Effective Mass Klein-Gordon Equation for Yukawa potential

C.A. Onate[1], A.N. Ikot[2], M.C. Onyeaju[2] and O. Ebomwonyi[3]
[1]*Department of Physical Sciences, Landmark University, Omu-Aran, Nigeria.*
[2]*Theoretical Physics Group, Physics Department, University of Port Harcourt, Nigeria.*
[3]*Physics Department, University of Benin, Benin City, Nigeria.*

## Abstract

The analytical solutions of the Klein-Gordon equation with the Yukawa potential is presented within the framework of an approximation to the centrifugal potential for any arbitrary $\ell$ state with the position-dependent mass using the parametric Nikiforov-Uvarov method. The energy eigenvalues and the corresponding wave function have been obtained. The energy for both the scalar potential and vector potential as well as the effect of the screening parameter on each of the energy for scalar potential and vector potential are investigated in detail. The nonrelativistic limit is obtained and numerical results are computed. It is found that our results for the constant mass and that of the nonrelativistic limit are in good agreement with the one in the literature.

Keywords: Wave equation; Klein Gordon equation; Position dependent mass; Yukawa potential, Centrifugal potential; Eigen solutions.

Pacs No: 03.65.Ge, 03.65.Ca, 03.65.Fd.

## Introduction

In the recent years, there has been an increasing interest in finding the analytical solutions of the relativistic Klein-Gordon equation [1-6]. This is because, the relativistic Klein-Gordon equation with some physical potentials of interest play significant roles in the relativistic quantum mechanics. Thus, some known physical potentials with the Klein-Gordon equation have received considerable attentions from many researchers in the past years. For instance, Ibrahim et al. [7], obtained analytical solutions of the N-dimensional Klein-Gordon equation and Dirac equation with Rosen-Morse potential. Falaye [8], solved the Klein-Gordon equation with ring-shaped potentials. Ikhdair and Hamzavi [9], investigated the effects of external fields on a two-dimensional Klein-Gordon equation of a particle under Pseudo-harmonic oscillator interaction. Hamzavi et al. [10] also studied spinless particles in the field of unequal scalar-vector Yukawa potentials. Dong [11], studied relativistic treatment of the spinless particles subject to a rotating Deng-Fan oscillator. Sun and Dong [12], studied relativistic treatment of the spinless particles subject to Tietz-Wei oscillator, Wei and Liu [13], obtained relativistic bound states of the hyperbolical potential with the centrifugal term. Ikhdair and Sever [14], obtained exact solutions of the Klein-Gordon equation for the PT-symmetric generalized woods-saxon potential using Nikiforov-Uvarov method. In addition, Ikhdair and Sever [15], also studied the exact bound states of the D-dimensional Klein-

Email:[1]oaclems14@physicist.net:

Gordon equation with equal scalar and vector ring-shaped pseudoharmonic potentials. Berkdemir [16], investigated the relativistic treatment of a spin-zero particle subject to the Kratzer-type potential. Dong et al. [17], studied the Klein-Gordon equation with a Coulomb potential in D dimensions. It is noted that some of the potentials mentioned above do not admit exact solutions due to the existence of the inverse squared term or centrifugal term attached to them. A typical example of such potentials is the Yukawa potential. The Yukawa potential was proposed by Hideki Yukawa [18] in the 1930s. This potential was showed to have arisen from the exchange of a massive scalar field such as the field of massive boson [18]. The Yukawa potential has various applications in the field of studied. The Yukawa potential modified for the finite size of the dispersing particle was represented with sufficient accuracy the influence of ionic concentration of the rigidity of colloidal systems of the polystyrene spheres and ovalbumin molecules [19, 20]. It was equally applied to the auto-controlled mechanism of the ovalbumin molecules in aqueous systems [21]. Matsumoto and Inoue [22] also used Yukawa potential to analyze the novel phenomenon in a solidlike structure in ovalbumin aqueous collides. The Yukawa potential is of the form [18]

$$V_Y(r) = -\frac{\lambda\, \mathrm{e}^{-\delta r}}{r}, \tag{1}$$

where $\lambda$ is the strength of the potential and $\delta$ is the screening parameter. When the screening parameter tends to zero, the Yukawa potential (1) reduces to pure Coulomb potential.

$$\lim_{\delta\to 0} V_Y(r) = -\frac{\lambda}{r}. \tag{2}$$

The Yukawa potential interacts with the Coulomb potential to generate the Hellmann potential [23] which has vigorous application in the representation of electron core interaction [24], alkali hydride molecules [25] and the study of the inner shell ionization problems [26]. The effective mass Klein-Gordon equation with Yukawa potential was studied by Arda and Sever [27] for a case when the scalar potential equals zero i.e. $S(r)=0$ with mass function $m_0$ and $m_1$. What is desired here is to investigate the analytical solutions of the effective mass Klein-Gordon equation with Yukawa potential for a case when $V(r)\neq S(r)$ and a case when $V(r)=0.$ The scheme of our work is as follow: In the next section, we present a brief parametric Nikiforov-Uvarov method. In section 3, the bound state solution of the effective mass Klein-Gordon equation is presented while in the last section, we give the concluding remark.

## 2. Parametric Nikiforov-Uvarov Method.

Given the general form of the Schrödinger equation as [28, 29],

$$U''(r) + \left[\frac{\alpha_1 - \alpha_2 s}{s(1-\alpha_3 s)}\right]U'(s) + \left[\frac{-\zeta_1 s^2 + \zeta_2 s - \zeta}{\left(s(1-s)\right)^2}\right]U(s) = 0. \tag{3}$$

Tezcan and Sever [29, 30] gave the energy condition as

$$n\alpha_2-(2n+1)\alpha_5+n(n-1)\alpha_3+\alpha_7+2\alpha_3\alpha_8+(2n+1)\sqrt{\alpha_9}=-\sqrt{\alpha_8}\left[\alpha_3(2n+1)+2\sqrt{\alpha_9}\right], \tag{4}$$

and the corresponding wave function is given as

$$U_{n,\ell}(s)=N_{n,\ell}s^{\alpha_{12}}(1-\alpha_3 s)^{-\alpha_{12}-\frac{\alpha_{13}}{\alpha_3}}P_n^{\left(\alpha_{10}-1,\frac{\alpha_{11}}{\alpha_3}-\alpha_{10}-1\right)}(1-2\alpha_3 s). \tag{5}$$

where $P_n^{(\mu,v)}(x),\ \mu>-1,\ v>-1,$ are Jacobi polynomials. The parameters $\alpha_i$ $(i=1,2,3,..........,13)$ are deduced as follows:

$$\left.\begin{array}{l}\alpha_4=\dfrac{1-\alpha_1}{2},\alpha_5\dfrac{\alpha_2-2\alpha_3}{2},\alpha_6=\alpha_5^2+\zeta_1,\alpha_7=2\alpha_4\alpha_5-\zeta_2,\\ \alpha_8=\alpha_4^2+\zeta_3,\alpha_9=\alpha_3(\alpha_7+\alpha_3\alpha_8)+\alpha_6,\alpha_{10}=\alpha_1+2\alpha_4+2\sqrt{\alpha_8},\\ \alpha_{11}=\alpha_2-2\alpha_5+2\left(\sqrt{\alpha_9}+\alpha_3\sqrt{\alpha_8}\right),\alpha_{12}=\alpha_4+\sqrt{\alpha_8},\\ \alpha_{13}=\alpha_5-\left(\sqrt{\alpha_9}+\alpha_5\sqrt{\alpha_8}\right).\end{array}\right\}. \tag{6}$$

**3. Bound state Solution**

The Klein-Gordon Equation with Scalar potential $S(r)$ and Vector Potential $V(r)$ of a particle of mass M and relativistic energy $E_{n\ell}$ in natural unit $(\hbar=c=1)$ is given as

$$\left[\frac{d^2}{dr^2}+(M+S(r))^2-(E_{n\ell}-V(r))^2-\frac{\ell(\ell+1)}{r^2}\right]U_{n\ell}(r)=0. \tag{7}$$

In order to solve Eq. (7) above, we take the mass function as

$$M(r)=m_0+\frac{m_1e^{-\delta r}}{1-e^{-\delta r}}, \tag{8}$$

where $m_0$ and $m_1$ are two positive constants. The scalar and vector potentials are taken as the Yukawa potential respectively:

$$S(r)=\frac{-S_0e^{-\delta r}}{r}, \tag{9}$$

$$V(r)=\frac{-V_0e^{-\delta r}}{r}. \tag{10}$$

Due to the presence of the centrifugal term in Eq. (7), we resort to apply the following approximation-type to deal with the centrifugal term

$$\frac{\ell(\ell+1)}{r^2}=\frac{\ell(\ell+1)\delta^2}{\left(1-e^{-\delta r}\right)^2}. \tag{11}$$

Now, substituting Eqs. (8), (9), (10) and (11) into Eq. (7) and by defining $y = e^{-\delta r}$ we have

$$\frac{d^2 U_{n\ell}(y)}{dy^2} + \frac{1-y}{y(1-y)} \frac{dU_{n\ell}(y)}{dy} + \frac{A_1 y^2 + A_2 s + A_3}{\left(y(1-y)\right)^2} U_{n\ell}(y) = 0, \tag{12}$$

where

$$A_1 = \frac{E_{n\ell}^2 - m_0^2 + 2m_0 m_1}{\delta^2} + \frac{2m_1 S_0 - 2m_0 S_0 - 2V_0 E_{n\ell}}{\delta} + V_0^2 - m_1^2 S_0^2, \tag{13}$$

$$A_2 = \frac{2E_{n\ell}^2 - 2m_0^2 + 2m_0 m_1}{\delta^2} + \frac{-2m_0 S_0 - 2V_0 E_{n\ell}}{\delta}, \tag{14}$$

$$A_3 = \frac{E_{n\ell}^2 - m_0^2}{\delta^2} + \ell(\ell+1). \tag{15}$$

Comparing Eq. (12) with Eq. (3), we deduce the analytic values given in the appendix. Using the analytic values in the appendix into Eq. (4), the energy equation and the corresponding wave function for the effective mass function are obtain respectively as

$$E_{n\ell}^2 = m_0^2 - \ell(\ell+1)\delta^2 - \delta^2 \left( \frac{\aleph - \frac{1}{2}\left(1 + (n+1)\sqrt{(1+2\ell)^2 + 4\left(V_0^2 - m_1^2 S_0^2\right) + \frac{8m_1 S_0}{\delta}}\right)}{2n + 1 + \sqrt{(1+2\ell)^2 + 4\left(V_0^2 - m_1^2 S_0^2\right) + \frac{8m_1 S_0}{\delta}}} \right)^2, \tag{16}$$

$$U_{n,\ell}(y) = N_{n,\ell}\, y^{\sqrt{A_3}} (1-y)^{-\frac{1}{2}-\frac{1}{2}\sqrt{(1+2\ell)^2 + \frac{8m_1 S_0}{\delta} + 4\left(V_0^2 - m_1^2 S_0^2\right)}}\, P_n^{\left(2\sqrt{A_3},\, \sqrt{(1+2\ell)^2 + \frac{8m_1 S_0}{\delta} + 4\left(V_0^2 - m_1^2 S_0^2\right)}\right)}(1-2y). \tag{17}$$

where

$$\aleph = -n(n+1) - 2\ell(\ell+1) + \frac{2m_0 S_0 + 2V_0 E_{n\ell}}{\delta} - \frac{2m_0 m_1}{\delta^2}. \tag{18}$$

**Case 2.**

When the scalar potential equal to zero i.e. $S(r) = 0,$ then, Eq. (7) reduces to

$$\left[\frac{d^2}{dr^2} + M^2 - \left(E_{n\ell} - V(r)\right)^2 - \frac{\ell(\ell+1)}{r^2}\right] U_{n\ell}(r) = 0, \tag{19}$$

and Eq. (16) becomes

$$E_{n\ell}^2 = m_0^2 - \ell(\ell+1)\delta^2 - \delta^2 \left( \frac{-n(n+1) - 2\ell(\ell+1) + \frac{2V_0 E_{n\ell}}{\delta} - \frac{2m_0 m_1}{\delta^2}. - \frac{1}{2}\left(1 + (n+1)\sqrt{(1+2\ell)^2 + 4V_0^2}\right)}{2n+1+\sqrt{(1+2\ell)^2 + 4V_0^2}} \right)^2 . \quad (20)$$

**Case 3.**

When the vector potential equal to zero i.e. $V(r) = 0,$ Eq. (7) reduces to

$$\left[ \frac{d^2}{dr^2} + (M + S(r))^2 - E_{n\ell}{}^2 - \frac{\ell(\ell+1)}{r^2} \right] U_{n\ell}(r) = 0, \quad (21)$$

and then, Eq. (16) reduces to

$$E_{n\ell}^2 = m_0^2 - \ell(\ell+1)\delta^2 - \delta^2 \left( \frac{\aleph_0 - \frac{1}{2}\left(1 + (n+1)\sqrt{(1+2\ell)^2 + 4m_1^2 S_0^2 + \frac{8m_1 S_0}{\delta}}\right)}{2n+1+\sqrt{(1+2\ell)^2 + 4m_1^2 S_0^2 + \frac{8m_1 S_0}{\delta}}} \right)^2 , \quad (22)$$

where

$$\aleph_0 = -n(n+1) - 2\ell(\ell+1) + \frac{2m_0 S_0}{\delta} - \frac{2m_0 m_1}{\delta^2}. \quad (23)$$

**Discussion**

In Table 1, we compared our results with the previous results. It is observed that our results show a good agreement with the previous one by Arda and Sever who obtained the solution of Klein-Gordon equation for a mass function with only vector Yukawa potential. In Table 2, we reported the eigenvalues with different values of the scalar and vector potentials. It is observed that the eigenvalues obtained with $S_0 > V_0$ are lesser than their counterpart with $V_0 > S_0$. In Table 3, we reported the eigenvalues for vector potential and scalar potential separately. It is noted from the Table that in the case of vector potential, the positive and negative energy eigenvalues obtained differs but for the scalar potential, the positive and negative energy eigenvalues are numerically the same. It is also deduced from Table 4, that for the same numerical value used for vector and scalar potentials, the energy eigenvalues for the vector potential are higher in magnitude compare to that obtained with the scalar potential. In Tables 4 (a) and (b), we presented the numerical results for a constant mass with $S_0 = V_0$, $S_0 < V_0$ and $S_0 > V_0$ respectively. In Tables 5 and 6, we computed numerical results for various states, angular momentum quantum number and the potential strength. These results are compared with results obtained from analytical method and AIM.

For a constant mass, $m_1 = 0$ and $m_0 = M$. Thus the energy equation of the Klein-Gordon equation with unequal scalar and vector potentials with constant mass becomes

$$E_{n\ell}^2 = M^2 - \ell(\ell+1)\delta^2 - \delta^2 \left( \frac{-n(n+1) + \dfrac{-2\left(\ell(\ell+1)\delta - MS_0 - V_0 E_{n\ell}\right)}{\delta} - \dfrac{1}{2}\left(1 + (n+1)\sqrt{(1+2\ell)^2 + 4V_0^2}\right)}{2n+1+\sqrt{(1+2\ell)^2 + 4V_0^2}} \right)^2 . \quad (24)$$

In a case where the scalar potential and the vector potential are equal for a constant mass, the energy equation becomes

$$E_{n\ell}^2 = M^2 - \ell(\ell+1)\delta^2 - \delta^2 \left( \frac{\dfrac{2\lambda(M+E_{n\ell})}{\delta} - \ell(\ell+1) - (1+\ell+n)^2}{2(n+\ell+1)} \right)^2 . \quad (25)$$

To the best of our knowledge, there is no experimental evidence for this solutions. Thus, our calculations are only of academic interest. It therefore becomes very necessary to test the accuracy of our calculations. To do this, we obtain the non-relativistic limit of Eq. (25) by considering the following transformation $M + E_{n\ell} = \frac{2\mu}{\hbar^2}$ and $M - E_{n\ell} = -E_{n\ell}$. The Klein-Gordon equation solved in this case is in a potential $2V$. However, Alhaidari et al. [31], pointed out that the Klein-Gordon equation whose non-relativistic limit equals the Schrödinger equation, is the Klein-Gordon equation with potential $V$ and not $2V$. The energy Eq. (25) is for Klein-Gordon equation with potential $2V$. For potential $V$, Eq. (25) becomes

$$E_{n\ell}^2 = M^2 - \ell(\ell+1)\delta^2 - \delta^2 \left( \frac{\dfrac{\lambda(M+E_{n\ell})}{\delta} - \ell(\ell+1) - (1+\ell+n)^2}{2(n+\ell+1)} \right)^2 . \quad (26)$$

Using the transformation given above, the non-relativistic limit of Eq. (26) becomes

$$E_{n\ell} = \frac{\ell(\ell+1)\delta^2\hbar^2}{2\mu} - \frac{\delta^2\hbar^2}{2\mu} \left( \frac{\dfrac{2\mu\lambda}{\delta\hbar^2} - \ell(\ell+1) - (1+\ell+n)^2}{2(n+\ell+1)} \right)^2 . \quad (27)$$

Table 1. Comparison of the eigenvalues for $S_0 = 0,\ M_0 = 1$ and $M_1 = 0.1$.

| | | | | $E_{n\ell}$ | | $-E_{n\ell}$ | |
|---|---|---|---|---|---|---|---|
| $n$ | $\ell$ | $\delta$ | $V_0$ | present | [27] | present | [27] |
| 0 | 0 | 0.01 | 0.10 | 0.999199 | 0.999181 | 0.998200 | 0.998173 |
| | | 0.10 | 0.01 | 0.994631 | 0.995475 | 0.995611 | 0.994475 |
| 1 | 0 | 0.01 | 0.10 | 0.999212 | 0.999987 | 0.998217 | 0.998985 |
| | | 0.10 | 0.01 | 0.995475 | 0.980294 | 0.994475 | 0.979294 |
| | 1 | 0.01 | 0.10 | 0.999789 | 0.999911 | 0.998569 | 0.998910 |
| | | 0.10 | 0.01 | 0.973433 | 0.954438 | 0.972211 | 0.953438 |
| 2 | 0 | 0.01 | 0.10 | 0.999834 | 0.999913 | 0.998836 | 0.998912 |
| | | 0.10 | 0.01 | 0.989183 | 0.954440 | 0.988183 | 0.953440 |
| | 1 | 0.01 | 0.10 | 0.999896 | 0.999622 | 0.998722 | 0.998622 |
| | | 0.10 | 0.01 | 0.964598 | 0.917015 | 0.963474 | 0.916015 |
| | 2 | 0.01 | 0.10 | 0.999639 | 0.999200 | 0.998401 | 0.998199 |
| | | 0.10 | 0.01 | 0.919255 | 0.866525 | 0.918015 | 0.865525 |

Table 2. Energy eigenvalues for $S_0 > V_0$ and $S_0 < V_0$ with $m_0 = 1$ and $m_1 = 0.1$.

| | | | $V_0 = 1,\ S_0 = 2.$ | | $V_0 = 2,\ S_0 = 1.$ | |
|---|---|---|---|---|---|---|
| $n$ | $\ell$ | $\delta$ | $E_{n\ell}$ | $-E_{n\ell}$ | $E_{n\ell}$ | $-E_{n\ell}$ |
| 0 | 0 | 0.1 | 0.7381321787 | 0.9987261947 | 0.8080766782 | 0.8542423602 |
| | | 0.2 | 0.3854265220 | 0.9386134340 | 0.6592013534 | 0.9237071838 |
| 1 | 0 | 0.1 | 0.8003546018 | 0.9920735842 | 0.8387227205 | 0.9387427245 |
| | | 0.2 | 0.5213698583 | 0.9079360473 | 0.6766325949 | 0.9896917707 |
| 0 | 1 | 0.1 | 0.7389521495 | 0.9828713267 | 0.7995194617 | 0.8954866609 |
| | | 0.2 | 0.3288767975 | 0.8271881889 | 0.5930853382 | 0.9369528056 |
| 1 | 1 | 0.1 | 0.7867708715 | 0.9734013059 | 0.8233389353 | 0.9541662973 |
| | | 0.2 | 0.4224599938 | 0.7928325762 | 0.5960377139 | 0.9591212145 |
| 0 | 2 | 0.1 | 0.7173741544 | 0.9441443364 | 0.7721856175 | 0.9322628032 |
| | | 0.2 | 0.0266460204 | 0.4791987774 | 0.4243296275 | 0.8669887330 |
| 2 | 0 | 0.1 | 0.8227535123 | 0.9815654305 | 0.8483420024 | 0.9793978934 |
| | | 0.2 | 0.5469616818 | 0.8606692272 | 0.6573090574 | 0.9980564492 |
| 2 | 1 | 0.1 | 0.8033280371 | 0.9608485139 | 0.8295093602 | 0.9808531902 |
| | | 0.2 | 0.4238361236 | 0.7330821942 | 0.5646457263 | 0.9389215255 |

Table 3. Energy eigenvalues for $S_0 = 0$ and $V_0 = 0$ with $m_0 = 1$ and $m_1 = 0.1$.

| | | | $V_0 = 1,\ S_0 = 0.$ | | $V_0 = 0,\ S_0 = 1.$ | |
|---|---|---|---|---|---|---|
| $n$ | $\ell$ | $\delta$ | $E_{n\ell}$ | $-E_{n\ell}$ | $E_{n\ell}$ | $-E_{n\ell}$ |
| 0 | 0 | 0.1 | 0.9987864821 | 0.4907214371 | 0.9987492177 | 0.9987492177 |
| | | 0.2 | 0.9305722548 | 0.7436217716 | 0.9121416177 | 0.9121416177 |
| 1 | 0 | 0.1 | 0.9939861024 | 0.8138776942 | 0.9931754873 | 0.9931754873 |
| | | 0.2 | 0.9228638305 | 0.9446130253 | 0.9099188673 | 0.9099188673 |
| 0 | 1 | 0.1 | 0.9815306962 | 0.7642101830 | 0.9816196551 | 0.9816196551 |
| | | 0.2 | 0.8578328241 | 0.8991725675 | 0.8237839752 | 0.8237839752 |
| 1 | 1 | 0.1 | 0.9747794364 | 0.9068297308 | 0.9737974815 | 0.9737974815 |
| | | 0.2 | 0.8397800592 | 0.9558052064 | 0.8110066978 | 0.8110066978 |
| 0 | 2 | 0.1 | 0.9441431506 | 0.9011913328 | 0.9423428588 | 0.9423428588 |
| | | 0.2 | 0.6884032443 | 0.8705342671 | 0.5996427023 | 0.5996427023 |
| 2 | 0 | 0.1 | 0.9865650821 | 0.9304278595 | 0.9845795617 | 0.9845795617 |
| | | 0.2 | 0.8961909438 | 0.9968308930 | 0.8814237163 | 0.8814237163 |
| 2 | 1 | 0.1 | 0.9654813434 | 0.9629901840 | 0.9633202543 | 0.9633202543 |
| | | 0.2 | 0.8033675472 | 0.9521219676 | 0.7715990292 | 0.7715990292 |

Table 4 (a): Eigenvalues $E_{n\ell}$ of the relativistic Klein-Gordon equation with constant mass.

| $n$ | $\ell$ | $\delta$ | $V_0 = S_0 = 2$ | $V_0 = 2,\ S_0 = 1,$ | $V_0 = 1,\ S_0 = 2.$ |
|---|---|---|---|---|---|
| 0 | 0 | 0.1 | 0.995866468 | 0.995866468 | 0.996475614 |
| | | 0.2 | 0.982824186 | 0.982824186 | 0.984881450 |
| 1 | 0 | 0.1 | 0.992499153 | 0.992499153 | 0.993782347 |
| | | 0.2 | 0.969268245 | 0.969268245 | 0.974184904 |
| 0 | 1 | 0.1 | 0.978373838 | 0.978373838 | 0.976895583 |
| | | 0.2 | 0.910805823 | 0.910805823 | 0.903633973 |
| 1 | 1 | 0.1 | 0.975350777 | 0.975350777 | 0.975371301 |
| | | 0.2 | 0.897867316 | 0.897867316 | 0.897189496 |
| 0 | 2 | 0.1 | 0.939659637 | 0.939659637 | 0.935630939 |
| | | 0.2 | 0.734690336 | 0.734690336 | 0.711036751 |
| 2 | 0 | 0.1 | 0.986828727 | 0.986828727 | 0.988499668 |
| | | 0.2 | 0.946076743 | 0.946076743 | 0.952602495 |
| 2 | 1 | 0.1 | 0.969455622 | 0.969455622 | 0.970089816 |
| | | 0.2 | 0.872287550 | 0.872287550 | 0.874197717 |

Table 4 (b): Eigenvalues $-E_{n\ell}$ of the relativistic Klein-Gordon equation with constant mass.

| $n$ | $\ell$ | $\delta$ | $V_0 = S_0 = 2$ | $V_0 = 2,\ S_0 = 1,$ | $V_0 = 1,\ S_0 = 2.$ |
|---|---|---|---|---|---|
| 0 | 0 | 0.1 | 0.999916580 | 0.999916580 | 0.999753637 |
| | | 0.2 | 0.999556933 | 0.999556933 | 0.998717146 |
| 1 | 0 | 0.1 | 0.997795164 | 0.997795164 | 0.997308924 |
| | | 0.2 | 0.990716697 | 0.990716697 | 0.988567173 |
| 0 | 1 | 0.1 | 0.986253971 | 0.986253971 | 0.982929931 |
| | | 0.2 | 0.942664230 | 0.942664230 | 0.928113741 |
| 1 | 1 | 0.1 | 0.982596249 | 0.982596249 | 0.980048101 |
| | | 0.2 | 0.927044373 | 0.927044373 | 0.916067025 |
| 0 | 2 | 0.1 | 0.951183898 | 0.951183898 | 0.934130919 |
| | | 0.2 | 0.780977831 | 0.780977831 | 0.741210646 |
| 2 | 0 | 0.1 | 0.992969952 | 0.992969952 | 0.992301070 |
| | | 0.2 | 0.970798163 | 0.970798163 | 0.967952069 |
| 2 | 1 | 0.1 | 0.976692728 | 0.976692728 | 0.974412564 |
| | | 0.2 | 0.901361032 | 0.901361032 | 0.891589676 |

Table 5: Comparison of energy eigenvalues $-E_{n\ell}$ of the Yukawa potential with $\hbar = \mu = 1,\ \lambda = \sqrt{2}$ and $\delta = \frac{g\lambda}{2}$ with other methods.

| State | $g$ | Present | [32] | [33] |
|---|---|---|---|---|
| 2p | 0.002 | 0.24700 | 0.24601 | 0.24601 |
| | 0.005 | 0.23802 | 0.24012 | 0.24010 |
| | 0.010 | 0.22905 | 0.23049 | 0.23040 |
| | 0.020 | 0.20820 | 0.21192 | 0.21160 |
| | 0.025 | 0.19632 | 0.20298 | 0.20250 |
| | 0.050 | 0.14622 | 0.16148 | 0.16000 |
| 3p | 0.002 | 0.10867 | 0.10716 | 0.10714 |
| | 0.005 | 0.10142 | 0.10141 | 0.10133 |
| | 0.010 | 0.09427 | 0.09230 | 0.09201 |
| | 0.020 | 0.07796 | 0.07570 | 0.07471 |
| | 0.025 | 0.06888 | 0.06815 | 0.06673 |
| | 0.050 | 0.03223 | 0.03711 | 0.03361 |
| 3d | 0.002 | 0.10778 | 0.10715 | 0.10714 |
| | 0.005 | 0.09779 | 0.10136 | 0.10133 |
| | 0.010 | 0.08783 | 0.09212 | 0.09201 |
| | 0.020 | 0.06464 | 0.07503 | 0.07471 |
| | 0.025 | 0.05144 | 0.06714 | 0.06673 |
| | 0.050 | 0.04331 | 0.03383 | 0.03361 |

Table 6: Comparison of energy eigenvalues $-E_{n\ell}$ of the Yukawa potential with $\hbar = 2\mu = 1,$ and $\delta = 0.4$ with other methods.

| $\lambda$ | $\ell$ | Present | [34] | [33] |
|---|---|---|---|---|
| 4 | 0 | 3.2400 | 3.2199 | 3.2400 |
| 8 | 0 | 14.440 | 14.420 | 14.440 |
| | 1 | 1.6400 | 2.4332 | 2.5600 |
| 16 | 0 | 60.840 | 60.819 | 60.840 |
| | 1 | 12.740 | 12.838 | 12.960 |
| 24 | 0 | 139.24 | 139.22 | 139.24 |
| | 1 | 30.840 | 31.239 | 31.360 |
| | 2 | 11.040 | 11.246 | 11.560 |

**Conclusion:**

In this work, we have examined the position dependent mass function of the Klein-Gordon equation with unequal scalar and vector Yukawa potential by employing a suitable approximation scheme to the centrifugal term in the framework of parametric Nikiforov-Uvarov method. The effect of both the scalar potential and vector potential are numerically studied in detail. By using some transformation, we have obtained the non-relativistic limit of the Klein-Gordon equation. Numerical results are computed for the non-relativistic limit and compared with results of other methods previously obtained. Our results are found to be in good agreement with the previous results.

**Appendix.**

Parametric constants

$$\alpha_1 = \alpha_2 = \alpha_3 = 1,$$

$$\alpha_4 = 0,\ \alpha_5 = -\frac{1}{2},$$

$$\alpha_6 = \frac{1}{4} + \frac{E_{n\ell}^2 - m_0^2 + 2m_0 m_1}{\delta^2} + \frac{2m_1 S_0 - 2m_0 S_0 - 2V_0 E_{n\ell}}{\delta} + V_0^2 - m_1^2 S_0^2,$$

$$\alpha_7 = -\left( \frac{2E_{n\ell}^2 - 2m_0^2 + 2m_0 m_1}{\delta^2} + \frac{-2m_0 S_0 - 2V_0 E_{n\ell}}{\delta} \right),$$

$$\alpha_8 = \frac{E_{n\ell}^2 - m_0^2}{\delta^2} + \ell(\ell+1).,$$

$$\alpha_9 = \frac{1}{4} + \frac{2m_1 S_0}{\delta} + \ell(\ell+1) + V_0^2 - m_1^2 S_0^2,$$

$$\alpha_{10} = 1 + 2\sqrt{\frac{E_{n\ell}^2 - m_0^2}{\delta^2} + \ell(\ell+1)},$$

$$\alpha_{11} = 2\left( 1 + \sqrt{\frac{E_{n\ell}^2 - m_0^2}{\delta^2} + \ell(\ell+1)} + \sqrt{\frac{1}{4} + \frac{2m_1 S_0}{\delta} + \ell(\ell+1) + V_0^2 - m_1^2 S_0^2.} \right),$$

$$\alpha_{12} = \sqrt{\frac{E_{n\ell}^2 - m_0^2}{\delta^2} + \ell(\ell+1)}$$

$$\alpha_{13} = -\frac{1}{2} - 1\left( \sqrt{\frac{E_{n\ell}^2 - m_0^2}{\delta^2} + \ell(\ell+1)} + \sqrt{\frac{1}{4} + \frac{2m_1 S_0}{\delta} + \ell(\ell+1) + V_0^2 - m_1^2 S_0^2.} \right).$$